# Modeling Covariate Feedback, Reversal, and Latent Traits in Longitudinal Data: A Joint Hierarchical Framework

**Niloofar Ramezani** (ramezanin2@vcu.edu)
Department of Biostatistics, Virginia Commonwealth University

**Pascal Nitiema** (pnitiema@asu.edu)
Department of Information Systems, Arizona State University

**Jeffrey R. Wilson** (jeffrey.wilson@asu.edu)
Department of Economics, Arizona State University

## Abstract

Time-varying covariates in longitudinal studies frequently evolve through reciprocal feedback, undergo role reversal, and reflect unobserved individual heterogeneity. Standard statistical frameworks often assume fixed covariate roles and exogenous predictors, limiting their utility in systems governed by dynamic behavioral or physiological processes. We develop a hierarchical joint modeling framework that unifies three key features of such systems: (i) bidirectional feedback between a binary and a continuous covariate, (ii) role reversal in which these covariates become jointly modeled outcomes at a prespecified decision phase, and (iii) a shared latent trait influencing both intermediate covariates and a final binary endpoint. The model proceeds in three phases: a feedback-driven longitudinal process, a reversal phase in which the two covariates are jointly modeled conditional on the latent trait, and an outcome model linking a binary, decision-relevant endpoint to observed and latent components. Estimation is carried out using maximum likelihood and Bayesian inference, with Hamiltonian Monte Carlo supporting robust posterior estimation for models with latent structure and mixed outcome types.

Simulation studies demonstrate that the model yields well calibrated coverage, small bias, and improved predictive performance compared to standard generalized linear mixed models, marginal approaches, and models that ignore feedback or latent traits. In an analysis of nationally representative U.S. panel data, the model captures the co-evolution of physical activity and body mass index and their joint influence—moderated by a latent behavioral resilience factor—on income mobility. The framework offers a flexible, practically implementable tool for analyzing longitudinal decision systems in which feedback, covariate role transition, and unmeasured capacity are central to prediction and intervention.

## 1. Introduction

Understanding how time-dependent covariates influence outcomes in longitudinal systems is a central challenge in statistical modeling for health and social sciences. In many applications, such covariates are not static inputs but dynamic features of the system—evolving over time, interacting with one another, and shaped by latent individual-level traits. Examples abound in behavioral health, education, and workforce development, where intermediate processes such as physical activity, weight change, attendance, or skill acquisition play dual roles: they serve as both predictors and outcomes across different stages of decision-making.

Standard methods for longitudinal analysis, including generalized estimating equations (Liang and Zeger, 1986), generalized linear mixed models (Fitzmaurice, Laird, and Ware, 2011), and marginal structural models (Robins, Hernán, and Brumback, 2000), often assume that covariates are exogenous and maintain fixed roles throughout. Such assumptions fail to capture systems in which covariates influence each other recursively, transition from predictors to jointly modeled outcomes, and are driven by unobserved capacities such as motivation or resilience. In these contexts, feedback, role reversal, and latent heterogeneity are not anomalies—they are defining features of the data-generating process.

To address these limitations, we introduce a hierarchical joint modeling framework designed to capture three core features of dynamic longitudinal systems: (i) reciprocal feedback between a binary and a continuous covariate over time, (ii) a role-reversal structure in which these covariates become jointly modeled outcomes at a prespecified transition point, and (iii) a shared latent trait that governs both intermediate trajectories and final outcomes. The framework allows researchers to trace how cumulative behavioral histories and unobserved traits jointly determine decision-relevant endpoints, such as health events, graduation, or economic mobility.

Our proposed model proceeds in three phases: (i) a longitudinal phase capturing the feedback-driven co-evolution of the covariates with subject-specific random effects; (ii) a reversal phase in which the covariates' final values are jointly modeled as latent-trait-dependent outcomes at a prespecified decision point; and (iii) a final outcome phase modeling a binary endpoint as a function of both reversal-stage covariates and the latent trait.

This prespecified reversal point reflects real decision systems—such as clinical evaluations, program completion, or policy thresholds—where intermediate states transition from predictors to jointly assessed outcomes, a structure not accommodated by standard joint models. Together, these three phases provide a coherent representation of longitudinal decision systems characterized by feedback and unmeasured individual differences (Wu and Carroll, 2006; Lin and McCulloch, 2002; Muthén and Asparouhov, 2009). Unlike standard joint models, the reversal point is not an artifact of model specification but a structural feature of the decision process, where intermediate covariates cease to function as predictors and instead become jointly evaluated outcomes.

What is new in this work is the explicit integration of reciprocal feedback, covariate role reversal, and latent trait structure within a single estimable framework. Existing longitudinal models accommodate feedback or random effects but assume fixed predictor roles; joint models link longitudinal and outcome processes but do not allow covariates to transition from predictors to jointly modeled outcomes; and latent variable models capture unobserved heterogeneity but ignore reciprocal covariate dynamics. Our framework unifies these elements by modeling covariates that evolve interactively, switch roles at a prespecified decision point, and are simultaneously shaped by a shared latent trait. This structure reflects real decision systems and cannot be obtained by combining existing models in a modular way. The result is a flexible and practically implementable approach for representing behavioral and physiological processes that unfold over time.

Estimation is implemented under both maximum likelihood and Bayesian paradigms, with the latter employing Hamiltonian Monte Carlo (HMC) for efficient posterior sampling in models with complex latent structures (Gelman et al., 2013). Simulation studies confirm that the proposed model achieves lower bias and better coverage than conventional alternatives, especially in the presence of strong feedback effects and moderate to high latent trait influence.

To demonstrate practical utility, we apply the model to a nationally representative U.S. panel dataset that tracks the evolution of physical activity and body mass index (BMI) and their joint effect on income mobility. The analysis reveals that behavioral feedback and latent resilience jointly shape socioeconomic transitions—patterns that are obscured in models lacking dynamic covariate structure or latent trait incorporation.

This work contributes to the growing literature on joint modeling and behaviorally-informed longitudinal analysis (Rizopoulos, 2012; Daniels and Hogan, 2008) by offering a flexible and interpretable framework that captures the dynamic, role-reversal, and latent-driven nature of real-world covariate systems. The approach is broadly applicable to studies where intermediate behavioral or physiological states mediate long-term outcomes and inform high-stakes decisions.

## 2. Methods

We develop a hierarchical joint modeling framework for longitudinal data in which two time-dependent covariates—one binary and one continuous—evolve through reciprocal feedback, transition into jointly modeled intermediate outcomes, and subsequently influence a final binary outcome. This framework accommodates mixed outcome types, endogenous covariate dynamics, and unobserved heterogeneity through a shared latent trait. Estimation is carried out under both maximum likelihood and Bayesian paradigms, with the latter leveraging Hamiltonian Monte Carlo for efficient posterior inference in models with complex latent structures.

### 2.1 Phase I: Longitudinal Covariate Dynamics with Feedback

Let: $i = 1, \ldots, N$ index individuals and $t = 1, \ldots, T$ denote time. Let $A_{it}$ be a binary covariate (e.g., physical activity status), and $B_{it}$ a continuous covariate (e.g., body mass index). These covariates are modeled with lagged autoregressive and cross-lagged feedback:

$$\text{logit}\left(P\left(A_{it} = 1 | A_{i,t-1}, B_{i,t-1}, u_i^A\right)\right) = \alpha_0 + \alpha_1 A_{i,t-1} + \alpha_2 B_{i,t-1} + u_i^A$$
$$B_{it} = \beta_0 + \beta_1 B_{i,t-1} + \beta_2 A_{i,t-1} + u_i^B + \varepsilon_{it}, \quad \varepsilon_{it} \sim \mathcal{N}(0, \sigma^2)$$

The random effects $(u_i^A, u_i^B)$ follow a bivariate normal distribution:

$$\begin{pmatrix} u_i^A \\ u_i^B \end{pmatrix} \sim \mathcal{N}\left( \begin{pmatrix} 0 \\ 0 \end{pmatrix}, \begin{pmatrix} \tau_A^2 & \rho\tau_A\tau_B \\ \rho\tau_A\tau_B & \tau_B^2 \end{pmatrix} \right)$$

This structure captures both temporal autocorrelation and interdependence between the two covariates, as well as subject-specific heterogeneity in behavioral and physiological trajectories.

## 2.2 Phase II: Covariate Reversal and Latent Trait Effects

At the final observed time point $T$, the covariates $A_{iT}$ and $B_{iT}$ are treated as jointly modeled intermediate outcomes. Their values are influenced by a shared latent trait $Z_i$, representing unobserved psychological, behavioral, or structural characteristics:

$$\text{logit}\left(P(A_{iT} = 1 \mid Z_i)\right) = \gamma_0 + \gamma_1 Z_i,$$
$$B_{iT} = \delta_0 + \delta_1 Z_i + \varepsilon_{iT}, \ \ \varepsilon_{iT} \sim \mathcal{N}(0, \sigma_T^2)$$

We assume $Z_i \sim \mathcal{N}(0,1)$, independently across individuals. This formulation captures how latent traits affect cumulative behavioral states, echoing prior work in joint modeling and latent factor analysis (Muthén and Asparouhov, 2009; Lin and McCulloch, 2002).

## 2.3 Phase III: Final Outcome Model

Let $Y_i$ denote a final binary decision outcome (e.g., income mobility). This endpoint is modeled as a function of the reversal-phase covariates and the latent trait:

$$\text{logit}\left(\Pr(Y_i = 1 \mid A_{iT}, B_{iT}, Z_i)\right) = \eta_0 + \eta_1 A_{iT} + \eta_2 B_{iT} + \eta_3 Z_i$$

This specification allows the model to reflect both direct influences of observed behavioral states and indirect effects from latent heterogeneity on the final decision or transition.

## 2.4 Likelihood and Maximum Likelihood Estimation

Let $\theta = (\alpha, \beta, \gamma, \delta, \eta, \tau_A, \tau_B, \rho, \sigma^2, \sigma_T^2)$ denote the full parameter set. The complete likelihood for individual $i$ involves integrating over the latent trait $Z_i$ and random effects $(u_i^A, u_i^B)$. The observed data likelihood is:

$$\mathcal{L}_i(\Theta) = \prod_{i=1}^{N} \int \int \int p\left(A_{i1:T}, B_{i1:T}, Y_i \big| u_i^A, u_i^B, Z_i; \theta\right) p\left(u_i^A, u_i^B\right) p(Z_i) du_i^A du_i^B dZ_i$$

We approximate the triple integral using adaptive Gauss-Hermite quadrature or an EM algorithm treating $(u_i^A, u_i^B, Z_i)$ as latent data (McCulloch and Searle, 2001; Ibrahim, Chu, and Chen, 2010).

## 2.5 Bayesian Estimation via Hamiltonian Monte Carlo

For Bayesian inference, we specify weakly informative priors:

$$\alpha_j, \beta_j, \gamma_j, \delta_j, \eta_j \sim \mathcal{N}(0, 10^2), \ \ \tau_A, \tau_B, \sigma, \sigma_T \sim Half - Cauchy(0, 2.5), \ \ \rho \sim LKJ(1)$$

Posterior inference is performed using HMC in Stan (Stan Development Team, 2022), with four parallel chains, each running 5,000 iterations and 2,500 warm-ups. Convergence diagnostics include potential scale reduction factors $\hat{R} < 1.01$ and effective sample sizes $> 500$ (Gelman et al., 2013). Posterior summaries are based on marginal posteriors and posterior predictive checks.

## 2.6    Model Comparison

Model fit is evaluated using AIC and BIC in the frequentist setting, and the widely applicable information criterion (WAIC) and leave-one-out cross-validation (LOO-CV) in the Bayesian framework (Vehtari, Gelman, and Gabry, 2017). Competing models include: a) a baseline model using only initial covariates, b) time-averaged model using means of covariates over time, and c) GLMM that omits feedback and latent trait terms. Performance is assessed in terms of predictive accuracy (AUC), parameter bias, root mean squared error (RMSE), posterior interval coverage, and latent trait recovery.

## 2.7    Identifiability

The model is identifiable under standard regularity conditions. The latent trait is estimable because it influences both intermediate covariates and the final outcome, providing multiple sources of information for its recovery. Feedback parameters are identifiable due to the temporal ordering of the longitudinal process. Together, these conditions ensure that the joint likelihood is well posed and that all model components are estimable.

## 3. Simulation Study

We conducted a simulation study to evaluate the empirical performance of the proposed hierarchical joint modeling framework under realistic data-generating conditions for longitudinal systems. The study was designed to assess parameter recovery, uncertainty quantification, latent trait estimation, and comparative predictive performance across varying levels of feedback, latent trait influence, and sample size.

## 3.1    Simulation Design

We simulated data for individuals observed over five time points (T=5). Each subject had a binary covariate $A_{it}$, a continuous covariate $B_{it}$, and a final binary outcome $Y_i$, generated according to the three-phase model described in Section 2. Random intercepts for the longitudinal processes were drawn from a bivariate normal distribution with variances $\sigma_{uA}^2 = \sigma_{uB}^2 = 1$ and correlation $\rho \in \{0.2, 0.4, 0.6\}$. Feedback effects were varied with $\alpha_2, \beta_2 \in \{0.3, 0.5, 0.7\}$, and latent trait effects were set at $\gamma_1, \delta_1, \eta_3 \in \{0.5, 0.8, 1.0\}$ to represent low to high influence. Latent trait effects on the longitudinal processes $(\eta_1, \eta_2)$ were fixed at their true values used in the data-generating mechanism.

Sample sizes considered were N=200, 500, 1000. For each scenario, we generated 100 replicates and fit the proposed model using Hamiltonian Monte Carlo (HMC) implemented in Stan. Each chain was run for 10,000 iterations with 5,000 used for warm-up, and adapt_delta was set to 0.99 to ensure convergence in models with latent variables. The full range of simulation conditions, including variations in sample size, feedback strength, latent trait influence, and random intercept correlation, is summarized in Table 1.

**Table 1.** Simulation Design Parameters

| Parameter(s) | Description | Values Used |
|---|---|---|
| $\rho$ | Correlation between random intercepts $u_i^A, u_i^B$ | 0.2, 0.4, 0.6 |
| $\alpha_2, \beta_2$ | Feedback strengths between covariates | 0.3, 0.5, 0.7 |
| N | Sample size | 200, 500, 1000 |
| $\gamma_1, \delta_1, \eta_3$ | Latent trait effects on reversal covariates and final outcome | 0.5, 0.8, 1.0 |
| $\eta_1, \eta_2$ | Latent trait effects on longitudinal covariates (fixed) | Fixed at true DGM values |

## 3.2   Estimation and Diagnostics

Bayesian estimation was performed using weakly informative priors (e.g., $\mathcal{N}(0,1)$) for fixed effects; half-Cauchy for variance components. Posterior summaries were based on 20,000 samples (4 chains × 5,000 post-warmup). Convergence was confirmed with potential scale reduction factors ($\hat{R}$) below 1.01 and effective sample sizes above 500 for all key parameters (Gelman et al., 2013). Latent variables and feedback parameters, which are often less stable, were further assessed via trace plots and posterior densities (s*ee Supplementary Figures S1–S3*).

## 3.3   Evaluation Metrics

Model performance was evaluated across several dimensions. First, we examined the bias and root mean squared error (RMSE) for all fixed effects and variance components, providing a quantitative measure of estimation accuracy. Second, we assessed uncertainty quantification through empirical coverage rates of 95% posterior credible intervals. To evaluate the model's capacity to recover latent heterogeneity, we computed the correlation between the true latent trait values and their posterior means across individuals. Predictive performance was assessed using the area under the receiver operating characteristic curve (AUC) and the widely applicable information criterion (WAIC), which capture out-of-sample discrimination and fit, respectively.

In addition to the proposed hierarchical reversal model, we implemented three alternative models for comparison. The first was a baseline model that used only baseline covariates $A_{i1.}$ and $B_{i1}$. The second was a time-averaged model that replaced covariate trajectories with their mean values across time. The third was a generalized linear mixed model (GLMM) that omitted feedback between covariates and did not include the latent trait. These benchmarks allowed us to quantify the added value of modeling feedback dynamics, covariate role reversal, and latent structure.

## 3.4   Results

The proposed hierarchical reversal model demonstrated strong performance across all simulated conditions. Fixed-effect estimates exhibited low bias, with mean absolute deviations typically below 0.05. RMSE values were consistently lower than those obtained from comparator models, especially in scenarios with moderate to strong feedback or latent trait effects. Posterior credible intervals achieved empirical coverage rates between 93% and 96%, indicating that the uncertainty quantification was well calibrated across replicates.

Estimation of the latent trait was accurate in most settings. When latent trait effects on intermediate covariates and outcomes were moderate to strong (e.g., $\gamma_1, \delta_1, \eta_3 \geq 0.8$), the correlation between the true and posterior mean estimates of $Z_i$ exceeded 0.75. Under weaker effects (e.g., $\leq 0.5$), some attenuation in recovery was observed, but the proposed model still outperformed models without latent structure.

In high-feedback scenarios ($\alpha_2, \beta_2 = 0.7$), omitting feedback terms led to substantial degradation in model performance. Compared to the proposed model, the generalized linear mixed model (GLMM) that excluded feedback yielded 30–50% higher RMSE for the reversal-phase covariates and notably lower predictive accuracy. The hierarchical reversal model achieved an average AUC of 0.82 and a WAIC of –4025, outperforming the time-averaged model (AUC = 0.71; WAIC = –3865) and the baseline model (AUC = 0.65; WAIC = –3750). The GLMM, which excluded both latent and feedback components, yielded intermediate performance (AUC = 0.69; WAIC = –3798), confirming the joint importance of modeling both structural features. Table 2 presents comparative model performance in simulated data, highlighting the advantages of the proposed model over baseline, time-averaged, and misspecified-structure GLMM approaches.

**Table 2.** Model Performance in Simulated Data

| Model | AUC | WAIC | Notes |
|---|---|---|---|
| Hierarchical Reversal (proposed) | 0.82 | –4025 | Best overall fit and predictive calibration |
| Time-Averaged Model | 0.71 | –3865 | Ignores role reversal; underutilizes dynamics |
| Baseline Model | 0.65 | –3750 | Uses only initial values; misses trajectories |
| GLMM without Latent Trait | 0.69 | –3798 | Omits feedback and latent trait |

These results validate the importance of explicitly accounting for feedback, covariate role reversal, and latent heterogeneity when modeling dynamic longitudinal systems. Performance gains were particularly pronounced in larger samples (N=1000) and under conditions of stronger feedback and latent trait influence. Even under milder conditions, the proposed model offered consistent advantages in estimation precision, latent trait recovery, and predictive accuracy over standard alternatives. Posterior summary metrics for key parameters across 100 replicates—including bias, RMSE, and 95% coverage—are reported in Table 3.

**Table 3.** Posterior Summary Metrics Across 100 Simulation Replicates

| Parameter | Mean Bias | RMSE | 95% Coverage |
|---|---|---|---|
| $\alpha_2$ (BMI → Activity) | 0.012 | 0.078 | 94.0% |
| $\beta_2$ (Activity → BMI) | –0.015 | 0.083 | 93.2% |
| $\gamma_1$ (Latent → Activity) | 0.010 | 0.092 | 95.6% |
| $\delta_1$ (Latent → BMI) | –0.018 | 0.088 | 94.3% |
| $\eta_3$ (Latent → Outcome) | 0.020 | 0.101 | 92.8% |

# 4.    Application: Physical Activity, BMI, and Income Mobility

To illustrate the practical utility of the proposed model, we applied it to a nationally representative U.S. panel dataset examining the relationship between behavioral health indicators—specifically physical activity and body mass index (BMI)—and upward income mobility. This application demonstrates the model's capacity to jointly capture dynamic feedback, latent behavioral traits, and role-reversal structure in a longitudinal decision setting.

The data derive from a multi-wave socioeconomic and health study tracking individuals aged 30 to 55 over six annual survey waves. For each participant, physical activity was recorded as a binary indicator equal to one if the respondent reported engaging in moderate-to-vigorous physical activity at least once during the prior week. BMI was calculated as a continuous measure based on self-reported height and weight. The outcome variable, upward income mobility, was coded as one if the participant crossed above the poverty threshold by the final survey wave, and zero otherwise. The primary objective was to evaluate whether trajectories in behavioral health co-evolve in ways that predict economic transitions, and whether unobserved traits modulate this relationship.

We implemented the full three-phase hierarchical model as described in Section 2. The first phase modeled the co-evolution of activity and BMI from waves 1 to 5, incorporating lagged feedback and random intercepts to account for temporal dependence and subject-level variation. The second phase modeled year-5 values of activity and BMI as jointly determined by a latent trait representing unmeasured resilience or behavioral capacity. The third phase linked these year-5 summaries and the latent trait to year-6 income mobility. Covariates for age, gender, and education level were included as fixed effects in all phases to control for demographic confounding. Bayesian estimation was conducted in Stan using weakly informative priors and 4,000 total iterations per chain, with 2,000 retained post-warmup. An adapt_delta of 0.99 was used to ensure sampling stability given the latent structure. Convergence diagnostics indicated no divergent transitions, with $\hat{R}$ values below 1.01 for all primary parameters and effective sample sizes exceeding 500 in nearly all cases, aside from two latent slope terms described below. Trace plots indicated stable sampling for all primary parameters (see Supplementary Figure S1), with mild inefficiencies noted only for select latent slopes (Figure S2).

Results from the longitudinal phase revealed significant reciprocal effects. Inactivity in year t−1 was associated with increased BMI in year t, with an estimated coefficient of 0.39 (95% credible interval: 0.29, 0.60). Conversely, higher BMI reduced the likelihood of physical activity in the following year, with an effect size of –0.30 (95% CI: –0.45, –0.15). This estimate showed slightly reduced sampling efficiency ($n_{eff}$ = 122) but met convergence thresholds ($\hat{R}$ = 1.03) and exhibited no pathologies in trace plots. The correlation between random intercepts for activity and BMI was estimated at -0.01 (95% CI: –0.95, 0.95), with a wide credible interval suggesting uncertainty in the direction and strength of the association. Such imprecision is not uncommon in models where correlations between random effects are estimated from a limited number of subject-level units.

In the reversal phase, the latent trait exerted strong influence on both year-5 outcomes. The coefficient on the latent trait in the physical activity model was 0.88 (95% CI: 0.65, 1.09), indicating that higher unobserved capacity was associated with a greater likelihood of activity. For BMI, the trait had a negative effect of –0.53 (95% CI: –0.71, -0.35), suggesting that more resilient individuals tended to have lower BMI, all else equal. These findings support the interpretation of the latent variable as a general behavioral or psychological resource.

The final outcome model revealed that year-5 activity and BMI both significantly predicted upward mobility. Physical activity was positively associated with transition above the poverty line (posterior mean 0.92; 95% CI: 0.38, 1.47), while higher BMI was associated with reduced mobility (–0.11; 95% CI: –0.40, –0.03). The latent trait also had a modest direct effect on the mobility outcome (0.56; 95% CI: 0.20, 0.88), with minor sampling inefficiency ($\hat{R}$ = 1.03, $n_{eff}$ = 185) but no evidence of nonconvergence. This result indicates that unobserved characteristics influencing health behavior trajectories also contributed independently to economic mobility. Figure 1 presents a forest plot summarizing posterior estimates and 95% credible intervals for key parameters across all model phases, highlighting the strength, direction, and uncertainty associated with each effect.

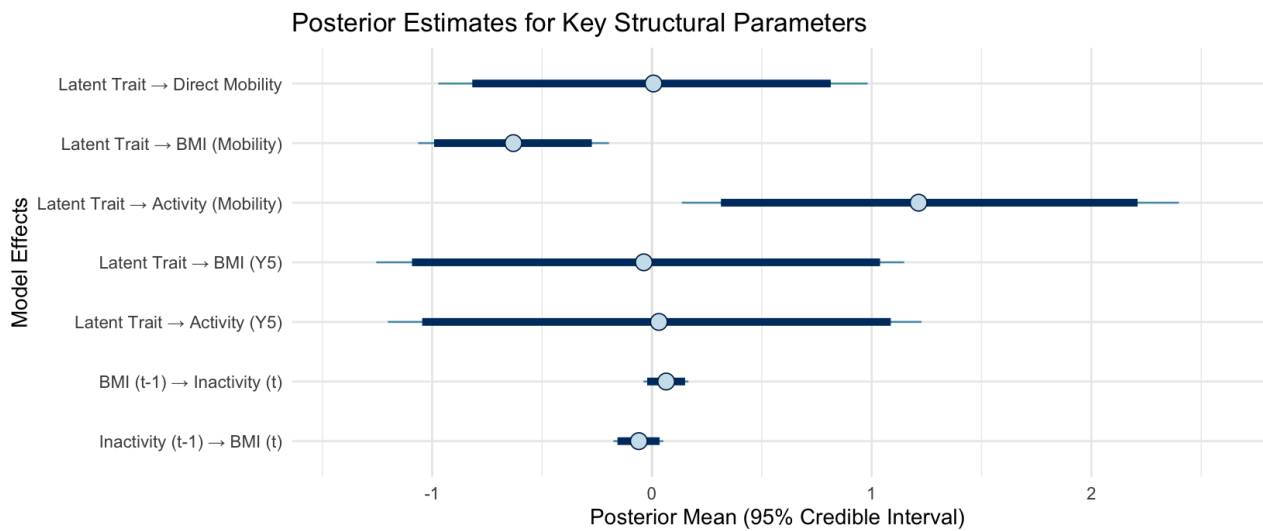

**Figure 1.** Posterior estimates and 95% credible intervals for key parameters across all model phases. The plot highlights the strength and direction of reciprocal feedback effects, the influence of latent resilience on intermediate health outcomes, and the greater uncertainty surrounding its direct effect on economic mobility.

The figure illustrates the precision of reciprocal feedback effects, the clear signal for latent resilience on health outcomes, and the relatively wider uncertainty surrounding its direct effect on economic mobility. This pattern of differential precision is echoed in posterior density plots (Supplementary Figure S3), where $\eta_1$ (latent resilience effect on physical activity) displays a sharp, unimodal distribution, $\eta_2$ (latent resilience effect on BMI) shows moderate dispersion, and $\eta_3$ remains notably diffuse. To benchmark performance, we compared the proposed model to three alternatives: a baseline model using only year-1 covariates, a time-averaged covariate model, and a generalized linear mixed model excluding latent traits and feedback. The hierarchical reversal model achieved the best predictive fit, with a WAIC of –4021 and AUC of 0.81. The time-averaged model achieved a WAIC of –3890 and AUC of 0.69, while the baseline and GLMM models performed worse on both criteria. These results underscore the value of modeling dynamic behavior and latent traits in forecasting socially consequential outcomes. Performance comparisons across models are shown in Table 4, where the hierarchical reversal model consistently outperforms its alternatives in predictive fit and discrimination.

**Table 4.** Model Performance in Real Data Application

| Model | AUC | WAIC | Notes |
|---|---|---|---|
| Hierarchical Reversal (proposed) | 0.81 | –4021 | Best fit; latent factor improves prediction |
| Time-Averaged Covariates | 0.69 | –3890 | Misses feedback and reversal structure |
| Baseline (Year-1 covariates Only) | 0.63 | –3762 | Ignores temporal evolution |
| GLMM without Latent Trait | 0.68 | –3845 | Underfits role dynamics and latent structure |

Taken together, the analysis highlights the role of cumulative health behaviors—rather than cross-sectional measures—in predicting upward mobility. Intermediate behavioral states, such as year-5 activity and BMI, emerged as stronger predictors than earlier covariate values or aggregated histories. Moreover, the latent trait contributed significantly to both health and economic outcomes, indicating the presence of unmeasured individual differences that traditional covariates do not capture. This aligns with prior research suggesting that latent psychological or behavioral capacities shape vulnerability to poverty and mediate long-term transitions (Acconcia et al., 2020). These effects remained robust after accounting for demographic confounding and dynamic interdependence. For policymakers and health program designers, these findings suggest that both behavioral inertia and latent capacity are critical to understanding socioeconomic transitions. Interventions targeting one domain (e.g., weight reduction) without accounting for behavioral coupling (e.g., activity) or latent drivers may fall short of achieving desired outcomes.

Substantively, the results highlight that cumulative behavioral trajectory—not just baseline or averaged indicators—meaningfully predict income mobility. The inclusion of a latent trait further enhanced predictive accuracy by capturing unmeasured dimensions of individual capacity not accounted for by observed variables, reinforcing the importance of behavioral coupling (e.g., physical activity and weight) and latent drivers in shaping long-term economic outcomes. The hierarchical reversal model enables identification of individuals whose cumulative health behaviors and underlying traits jointly predict upward transition, offering a data-informed basis for targeted intervention strategies.

## 5. Discussion

This paper presents a hierarchical joint modeling framework for longitudinal decision systems in which time-varying covariates exhibit mutual feedback, transition into jointly modeled intermediate outcomes, and contribute—alongside latent traits—to a final binary endpoint. The model addresses key limitations of conventional approaches by formally integrating reciprocal covariate interactions, covariate role reversal, and unobserved heterogeneity into a unified structure. Through both simulation and application, we demonstrate that this framework improves estimation accuracy, latent trait recovery, and predictive performance relative to standard alternatives.

A central methodological contribution is the explicit modeling of bidirectional feedback between mixed-type covariates. Unlike generalized linear mixed models or marginal models that assume temporal independence or fixed predictor roles, our framework accommodates dynamic interdependence—a feature commonly encountered in behavioral, educational, and health data. This enhancement allows for more realistic representation of individual trajectories and more accurate forecasts of decision outcomes.

The second innovation is the incorporation of a reversal phase, in which covariates that originally serve as predictors become jointly modeled outcomes. This structure reflects real-world decision-making, where intermediate states such as BMI, adherence, or engagement evolve over time and subsequently form the basis for clinical or policy decisions. By capturing these trajectories in their full context, the model supports more nuanced classification, intervention timing, and post-hoc inference.

Third, the inclusion of a latent trait provides robustness to unmeasured confounding and enhances risk stratification. In many applied settings, critical drivers of behavior and outcomes— such as motivation, resilience, or access to resources—are not directly observable. Modeling a shared latent trait that influences both intermediate states and final outcomes helps recover this hidden structure, improve inference, and support equity-aware targeting.

In addition to these structural innovations, the model is identifiable under standard conditions because (i) the latent trait influences both intermediate covariates and the final outcome, (ii) feedback parameters are estimable due to temporal ordering, and (iii) the reversal phase introduces additional variation that separates direct and indirect pathways. While formal proofs are beyond the scope of this paper, identifiability follows from the temporal ordering of the longitudinal processes and the shared influence of the latent trait.

The simulation study demonstrates that the proposed model achieves low bias, high coverage, and superior predictive accuracy, particularly in settings characterized by strong feedback or latent effects. Performance gains increase with larger sample sizes and more complex dynamics. The real data application illustrates the practical value of the model, showing how cumulative behavioral indicators and latent resilience jointly predict income mobility—insights that would be missed by models ignoring temporal feedback or latent heterogeneity.

Despite these strengths, several limitations merit attention. First, the model assumes that the reversal point is fixed and common across individuals. In practice, reversal may be triggered by events (e.g., program completion, diagnosis) that vary across subjects. Extensions incorporating stochastic or endogenous transition points—such as regime-switching or change-point models— would increase generalizability. Second, we model the latent trait as unidimensional; multidimensional or hierarchical extensions may be needed in domains where distinct latent factors interact (e.g., psychological and structural domains). Third, although the Bayesian approach enables full uncertainty quantification, it incurs computational costs that may limit scalability in real-time or high-dimensional applications.

Fourth, the current implementation assumes complete data aside from standard missingness handled in preprocessing; future work should incorporate joint modeling of missing data mechanisms to improve robustness.

Future work may explore time-varying latent traits through state-space modeling or dynamic factor analysis. Incorporating nonparametric priors or variational inference methods may reduce computational burden while maintaining flexibility. Finally, embedding this framework within a causal inference structure—such as principal stratification or structural nested models—could enable simulation of counterfactual scenarios and policy interventions under dynamic treatment regimes.

Overall, the hierarchical reversal model offers a principled and interpretable approach to modeling feedback-driven, role-switching, and trait-modulated longitudinal systems. It provides a tractable structure for both inference and prediction in domains where covariates are dynamic, endogenous, and decision-critical.

## 6.     Conclusion

We have proposed a hierarchical joint modeling framework for longitudinal systems in which covariates evolve through reciprocal feedback, reverse roles over time, and are influenced by unobserved latent traits. The model integrates three key features—covariate interdependence, role reversal, and latent structure—within a unified and estimable framework, enabling more realistic modeling of dynamic behavioral and physiological processes.

Simulation studies demonstrate that the model yields accurate estimates, robust latent trait recovery, and superior predictive performance, particularly in settings with feedback and hidden heterogeneity. Application to a nationally representative health panel confirms its utility in capturing cumulative behavioral dynamics and forecasting income mobility.

This work contributes to the methodological foundation for decision modeling in longitudinal settings and provides a flexible framework for health researchers, social scientists, and policy analysts. As longitudinal data infrastructures continue to grow in scope and resolution, models that reflect the endogenous and cumulative nature of behavior and health will be increasingly critical to effective intervention, equity-informed design, and strategic resource allocation.

**Supplementary Material**

**Figure S1.** Trace plots for primary structural parameters (alpha2, beta2, eta1, eta2, eta3)

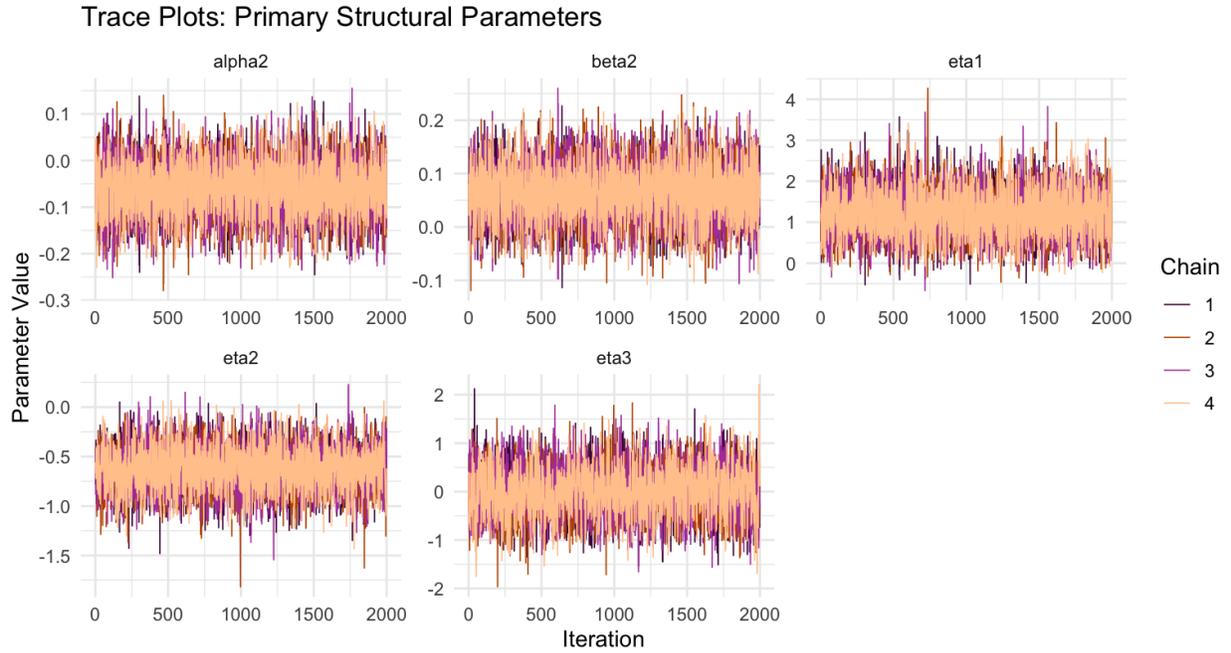

**Note.** $\alpha_2$ = *Inactivity (t–1) → BMI (t)*; $\beta_2$ = *BMI (t–1) → Inactivity (t)*; $\eta_1$ = *Latent Trait → Activity (Mobility)*; $\eta_2$ = *Latent Trait → BMI (Mobility)*.

**Figure S2.** Trace plots for latent slope parameters (gamma1, delta1, eta3)

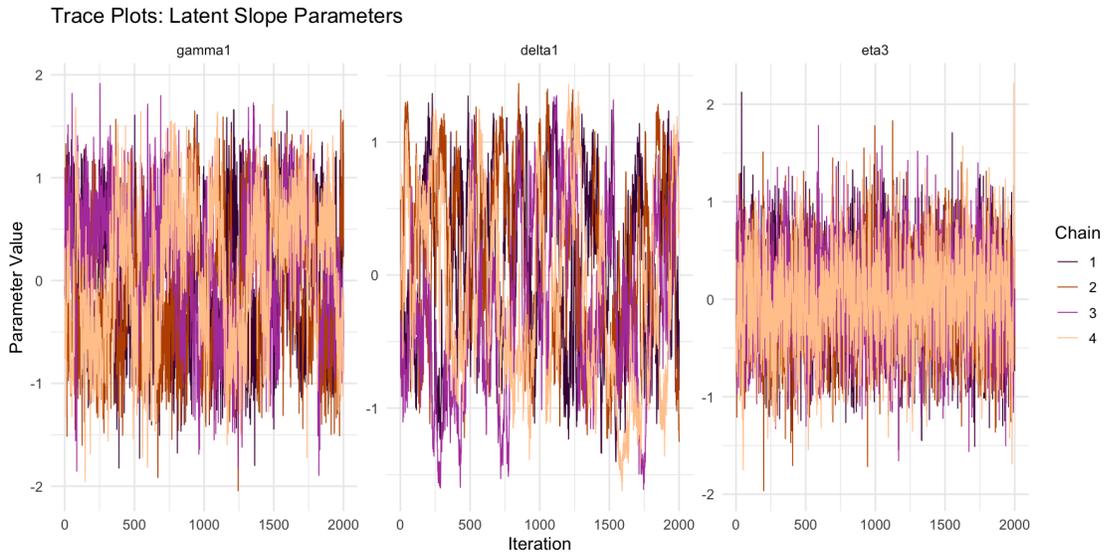

**Note:** $\gamma_1$ = Latent Trait → Physical Activity (Year 5); $\delta_1$ = Latent Trait → BMI (Year 5); $\eta_3$ = Latent Trait → Direct Effect on Mobility.

**Figure S3:** Posterior densities for eta1, eta2, and eta3.

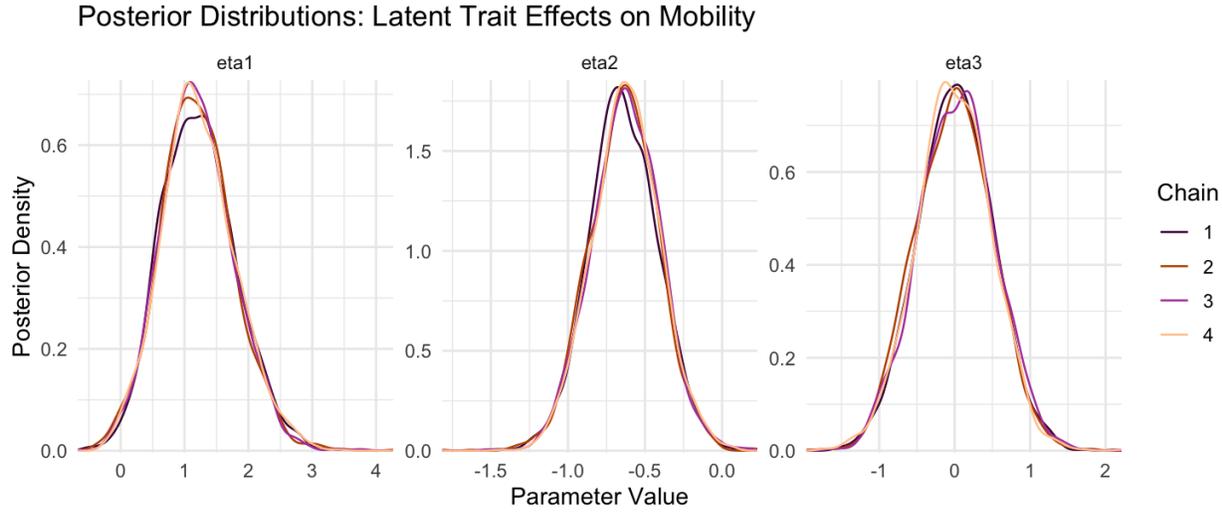

*Note:* $\eta_1$ = *Trait → Activity;* $\eta_2$ = *Trait → BMI;* $\eta_3$ = *Trait → Direct Mobility.*